\documentclass{aastex}
\usepackage{spr-astr-addons}
\usepackage{url}

\RequirePackage{color}
\def\ltsima{$\; \buildrel < \over \sim \;$}
\def\simlt{\lower.5ex\hbox{\ltsima}}
\def\gtsima{$\; \buildrel > \over \sim \;$}
\def\simgt{\lower.5ex\hbox{\gtsima}}        
\def\kms{km~s$^{-1}$}

\shorttitle{Gravitational redshift in optical AGN spectra}
\shortauthors{Bon et al.}

\begin{document}


\title{GRAVITATIONAL REDSHIFT OF EMISSION LINES IN THE AGN SPECTRA}


\author{Nata\v sa Bon\altaffilmark{1}, Edi
Bon\altaffilmark{1},  Paola Marziani\altaffilmark{2} and Predrag 
Jovanovi\'c\altaffilmark{1}}
\affil{$^1$ Astronomical Observatory, Belgrade, Serbia}
\affil{$^2$ INAF, Padova, Italy}
\email{nbon@aob.bg.ac.rs}



\begin{abstract}

The detection of  gravitationally redshifted optical emission lines has been
reported just for a few active galaxies. In this paper we give a short overview
of 
studies that analyzed or exploited the detection of the gravitational redshift
in
optical AGN spectra.
In addition, we tested the consistency of gravitational  redshift as the
physical 
origin of the redward shifts observed in their spectra using a sample of
$\approx$ 50  Hamburg-ESO intermediate to high redshift quasars that are among
the most luminous quasars known ($10^{47}  \simlt L \simlt 10^{48}$ erg/s), and
are expected to host very massive black holes.  To this aim we modeled the line
profile with accretion disk models around a black hole. 

\end{abstract}






\section{Introduction}

\indent A large fraction  of active galactic nuclei show
red asymmetries in their broad optical emission lines. The asymmetry is strong
in 
Balmer lines, but even stronger redward asymmetry is displayed in the lines that
originate closer to the central black hole (BH) such as HeII$\lambda$4686,
HeI$\lambda$5876 or CIV $\lambda$1549 \citep{Peterson99,Peterson01,Koll03}.
One of the proposed explanation of the
red asymmetry is that the broad emission line gas is gravitationally dominated
by the central black hole. 
 In this regard, following the dichotomy of quasars proposed by
\cite{Sulentic00} onto Pop. A and Pop B. according to their spectral properties,
a high fraction of Pop. B objects that are associated to AGN
with the most massive BHs \citep{Zamfir10} show redshifted profiles. 
\cite{Corbin95} gave the quantitative test if
the gravitational redshift ($\rm z_{grav}$) model is possible, by calculating
the mass of the central source, and found that the mass
corresponding to the shift $z \sim$0.001 should yield to the mass of
around $10^9 M_{\odot}$. This mass  is close to the largest masses 
estimated for quasar black hole, since they rarely exceed $5 \cdot 10^{9}$ 
$M_{\odot}$, 
and are almost always $ \lesssim 10^{10}  M_{\odot}$
\citep{Czerny04,Sul06,MarzSul12}. \\

\indent In this paper we give a brief overview on previous gravitational
redshift analyses. Also, we consider the origin  of the redshift in the sample
of $\approx$50 Hamburg-ESO (HE) intermediate to high
redshift quasars, that are among the most luminous quasars known ($10^{47} 
\simlt L \simlt 10^{48}$ erg/s), and are expected to host very massive black
holes. We then matched simulated relativistic emission line profiles
obtained by ray tracing method with the optical emission H$\beta$
line profiles from HE sample, for the cases where centroids of line widths
measured on 1/4 of maximum line intensity were significantly shifted to the
red,  in order to see if their shapes could be distorted by
gravitational redshift. \\

\indent The paper is organized as follows. In Section 2
an overview on the gravitational redshift hypothesis is presented. Implications
of 
gravitational redshift   in the case of weak gravitational field are analyzed in
Section 3. In Section 4 we consider
the possibility that the gravitational redshift shapes the red wing of H$\beta$
profile.  Section 5 gives our conclusions on the gravitational redshift search
in 
optical AGN spectra.

\section{Overview: AGN emission lines and gravitational redshift}

The concept that gravitational redshift may influence quasar optical emission
lines
causing line  asymmetries was proposed by \cite{Netzer77}.\footnote{In the
following line
shifts will be considered with respect to the quasar rest frame i.e., after
removing the effect of cosmological redshift.}  There were some earlier attempts
 to consider gravitational redshift 
as an explanation for the cosmological redshift of quasars but this led to
a series of inconsistencies
\cite[see][]{Greenstein64}. 
 The systematic gravitational redshift may have a non-negligible effect
on the shape of spectral lines, when the emission region is in the gravitational
field of a massive black hole ($ M \sim 10^{7}-10^{10} M_{\odot}$)
\citep{Netzer77,Anderson81,Popovic95,Corbin97}. This leads to an asymmetric and
redshifted profile of emission lines, especially in broad lines (BL) or
their
very broad line (VBL) base that originate at distances of only few light days
from the central black hole \citep{Popovic95}.\\
\indent
  \citet{Peterson85} suggested
gravitational redshift as an explanation for the red shifted very broad wings 
in the optical emission line profiles  
in order to explain the redward asymmetric broad H$\beta$
profile of Arakelian 120, under the  assumption of very compact broad line
region (BLR). 
This effect was  more systematically investigated by \cite{Sulentic89} 
and \cite{ZS90} who assumed a single component model of the broad
line region, and by \cite{Corbin95} who assumed a two component model, a
very broad line region (VBLR, where profile bases are formed) and an
intermediate
line region (ILR, forming the profile cores) for the BLR.  
 \cite{Corbin97} studied relativistic effects on the emission from
the BLR assuming orbits for the emitting clouds in a Schwarzschild
geometry, and proposed a model for the BLR in which he distinguished two
emitting 
regions, the ILR and the VBLR, taking into account the relativistic effects on
the emission line profiles. In that paper he  suggested that redward profile
asymmetries can be produced by the effect of gravitational redshift on the
emission from the VBLR, when the VBLR is represented by a flattened ensemble of
clouds viewed nearly face-on, at a mean distance of few tens of gravitational
radii from the black hole. The redshift of the ILR component could be as
well explained with the same geometry of a flattened ensemble of clouds lying on
a distance of several
hundreds gravitational radii from the central black hole.\\

\citet{Corbin93,Corbin95} found that the red asymmetry of H$\beta$ line
increases with the increasing soft X-ray luminosity. This correlation may be
interpreted as a
progressive redshift of the very broad line region velocity centroid ($\sim$
5000-20000 \kms) with respect to the systemic velocity with increasing X-ray
luminosity. Also, \citet{Corbin95} noticed that, on average, radio-loud (RL)
objects
show stronger redward asymmetric profiles than radio-quiet (RQ) object of
comparable optical
luminosity. Under the gravitational redshift hypothesis, these differences could
be
explained as the larger masses of the black hole and the host galaxy for RL 
objects. This is consistent with the evidence that they are
merger products \citep{Schweizer82, KormendyDjorgovski89, Heckman86,
Smith86, Hutchings87, WillsonColbert95}. However, this interpretation may
not be correct because RQ in the same black hole mass domain are also found
\citep{Marz03, Zamfir10}. \\ 

\indent Since the variable part of the spectral line originates far
more closely to the central nuclei than the constant line
components, \cite{Peterson99, Koll97} and \cite{Koll03} assumed that the
gravitational
redshift effect might be present in this variable component. Therefore, they
used
the rms profile of the spectra from the monitoring campaign of NGC5548
\citep{Peterson99}, NGC4593 \citep{Koll97} and Mrk 110 \citep{Koll03},
measured the shift between the rms profile
and the mean profile, and identified the differential shift as gravitational
redshift.
\cite{Koll03} analyzed the shift in four emission lines H$\alpha$, H$\beta$,
HeI$\lambda$5876 and HeII$\lambda$4686 and found the strongest
$\rm z_{grav}$
in the HeII$\lambda$4686, as it originates closest to the black hole.
The confirmation that the optical emission lines in Mrk 110 are gravitationally
affected was given by \cite{MW06}, where emission lines are modeled under the
assumption that they originate in the vicinity of a black hole.\\

\subsection{Exploitation of gravitational redshift}

\indent  If line redshift with respect to rest frame  and redward line
asymmetries are indeed of gravitational origin, it would be possible to derive
parameters that are important in 
studies of AGN spectra, such as:
\begin{itemize}
 \item \textit{the SMBH
mass, with an orientation-independent estimate} - Knowing the gravitational
redshift $z_{grav}$, 
one can estimate an upper limit of
the central mass as $M_{grav} \sim c^{2} z_{grav} r G^{-1}$ \citep{ZS90}.
This
way of estimating the SMBH mass is independent of the geometry and inclination
of the object, but still requires the knowledge of the emitting region radius
(r).
Other methods for mass estimations that assume the virial motion
of the gas, such as $M_{orbital}=fv^{2}G^{-1}r$ \citep{Wandel99}, include the
dimensionless
factor $f$ that depends, among other things, on the geometry and orientation of
the BLR.
\item \textit{the inclination angle of the accretion disk} - in the case
of Mrk 110, \cite{Koll03} assumed the accretion disk
model of the BLR. Under this assumption, the comparison between $M_{grav}$ and 
$M_{orbital}$ give an estimation of the orientation angle $i$ of the accretion
disk : $M_{orbital}$/$M_{grav}$=$sin^{2}i$. Therefore, this method makes it
possible to estimate the inclination angle of the accretion disk for radio-quiet
AGNs. Until now, the inclination angle has been estimated only for a relatively
small number of radio-loud sources in which superluminal motion has been
detected \citep{Sulentic03,Rokaki03}. Recently, \citep{Piotrovic2015} presented
the new method for the estimations of the inclination of the broad emission
line region (BLR), based on the observations in a polarised light, under the
assumption of a flattened configuration of BLR, but this interesting method
has been tentatively applied until now only to a few tens of Seyfert nuclei.\\

\item \textit{Spin orientation of the supermassive black hole} - 
\cite{Koll03} went one step further in the interpretation of the
inclination angle. From the cross-correlation analysis he determined the
distances of the line emitting regions, using the rms profile.
He found that the variable HeII$\lambda4686$ line component originates at
a mean distance of 3.9 light days, that mean 230
Schwarzschild radii only.
Therefore, he suggested that the orientation axis of the inner accretion disk
is oriented parallel to the spin axis of the central black hole. 
\end{itemize}

\indent However, the nature of the systematic line shift still remains  an
open question, since shifts in both directions - redward, but also blueward are
observed. 
Redshifts of H$\beta$ are observed more frequently than blueshifts, and that
could indicate a systematic trend that may not be immediately explainable by
Doppler effect assuming plausible kinematic scenarios. 
\cite{SuzyC06} analysed symmetric aspect of H$\beta$ line and concluded
that H$\beta$ line profile is primarily sensitive to Eddington ratio, but that
inclination effects play role in some cases, such as narrow line Seyfert 1
galaxies (NLS1).
The fraction of
redshifted profiles is very high in Pop. B, usually
associated to AGN with the most massive BHs \citep{Zamfir10}. Beside
gravitational redshift, both radial and
transverse velocity could account for the observed term. Among them,
the gravitational term could be the most probable cause, since the
transverse
term
is on the same order of magnitude as the gravitational one, and in fact
implies a dominant gravitational field \citep{ZS90}.\\ 

\begin{table}[t]
\begin{center}
\scriptsize
\caption{Table shows the difference of the gravitational red shift
contribution calculated using Schwarzschild and weak field approximation.}
\label{tbl-1}
\begin{tabular}{@{}llll}
\tableline
Radius    &$\rm z_{grav}$\ & $\rm z_{grav}$\ & Diff.   \\
($\rm R_\mathrm{S}$) & Schwarzschild & Weak field   & \\
\tableline
20	&	0.025978 	&	0.025000	&	4\%	\\
30	&	0.017095 	&	0.016667	&	3\%	\\
40	&	0.012739 	&	0.012500	&	2\%	\\
100	&	0.005038	&	0.005000	&	1\%	\\
200	&	0.002509 	&	0.002500	&	0\%	\\
\tableline
\end{tabular}
\end{center}
\end{table}

\section{The test for the redshift origin of optical emission
lines in AGN sample - is it really gravitational redshift?}

A sample of $\approx$ 50  Hamburg-ESO (HE) intermediate to high redshift quasars
have been observed 
in the H$\beta$ spectral range using the IR spectrometer ISAAC
\citep[M09, and references therein]{Marz09}. 
These sources are among the most luminous quasars known ($10^{47}  \simlt L
\simlt 10^{48}$ erg/s), and are expected 
to host very massive black holes, as indicated by the large H$\beta$\ FWHM: all 
black hole masses  
computed from  the scaling relation of Vestergaard and Peterson (2006) are
indeed       
$\log M  >  9$. The Population B part of this sample  has a high prevalence of
redshifted 
centroids\footnote{Centroid is calculated as follows:
$\cal{C}\rm
(\dfrac{i}{4})=\frac{\lambda_{B}(\dfrac{i}{4})+\lambda_{R}(\dfrac{i}{4}) } { 2
}-\lambda_{0}$, for i=1,3, where $\lambda_{R}$, and $\lambda_{B}$, represent
the wavelength on the red and blue side, respectively, measured at $i/4$ level
of the line intensity.}:
$\approx$ 80\%\ of all Pop. B sources show a significant centroid
displacement to the 
red at 1/4 fractional intensity (Fig. 9  of M09). The H$\beta$  redward
asymmetry is well correlated with the black hole mass (Fig. 16  of M09). 

Therefore, these luminous sources represent an appropriate sample to test the
consistency of gravitational 
redshift as the physical origin of the redward shifts observed in Pop. B. Under
the assumption of weak field in the Schwarzschild metric, 
one can write for the combined effect of gravitational plus transverse redshift

\begin{equation}
\Delta z \approx   \frac{3}{2} \frac{G M}{R  c^2} =  \frac{3}{4}
\frac{R_{\mathrm{S}}}{R}
\end{equation}

If  $\cal{C}$ indicates the line centroid, then
$\Delta z = \frac{   {\cal C}}{c}$, and $R =  \frac{3 c }{4} {\cal C}^{-1}
R_{\mathrm{S}}$, or
$R \approx    225  \cdot {\cal C}_{1000}^{-1} ~~R_{\mathrm{S}}$, where the
centroid is now in units of 1000 \kms, and $R_\mathrm{S}$\ is the Schwarzschild
radius. Basically a centroid displacement of 1000 \kms\  corresponds to a
distance from the central mass of  about 200 $R_{\mathrm{S}}$.  We can consider
the centroid at 1/4 intensity ${\cal C}(\frac{1}{4}$), or after a  decomposition
of the H$\beta$\ profile into a broad component (BC) and a very broad
component (VBC), the peak of the VBC. For the HE
sample, the  median value of the VBC peak shifts is $\approx 2200$ \kms, and 
${\cal C}(\frac{1}{4}$) $\approx 1600$ \kms,  corresponding to just $\sim$ 100
Schwarzschild radii (note that even at about 30 $R_{S}$
 the weak field formula  differs by only 3\%\ from the correct Schwarzschild
expression, see Table \ref{tbl-1}). \\
\indent In Table \ref{tbl-2} we report several parameters for a sample of 16 HE
quasars (out of 28 of the sample of Sulentic et al., in preparation) showing an
H$\beta$\ redward asymmetry. The Columns of Table \ref{tbl-2} list in the
following order - bolometric luminosity, black hole mass, ${\cal
C}(\frac{1}{4}$),
VBC peak shift, VBC FWHM, fraction of flux in the VBC, the VBC luminosity, as
well as the distances needed for explaining the observed redward displacement
due to gravitational+transverse redshift, as derived from the  ${\cal
C}(\frac{1}{4}$) and from the VBC shift. 
 From the median mass
 $\log M \approx 9.65$, we derive   median distances of 140 (from ${\cal
C}(\frac{1}{4}$)) and 100 $R_{S}$\ (from VBC shift), which correspond to  linear
sizes of $\sim 1.8$\ and $1.4 \cdot 10^{17}$\ cm  (see Table \ref{tbl-2}). 
This radius may be considered  as an emissivity weighted radius, and certainly
not 
the inner radius of the disk.

Here we try to answer the question: Is there enough gas close to the central
black hole that is able to emit the very luminous redshifted emission components
observed in the HE quasars?
A first hint, at the expected H$\beta$ luminosity, may come from the assumption
of a  photoionized medium in case B. Under this assumption the covering factor
of the
emitting gas  should be less than 1 (typically 0.1 -- 0.2, no more than 0.5) for
gravitational redshift to produce a consistent result.   A low covering factor
is possible only for  high density ($\simgt 10^{11} $\ cm$^{-3}$) and large
column density
($\sim 10^{25}$\  cm$^{-2}$). However, at such  high $L$ and small radii, a
case B photoionization may  not be possible at the illuminated face of the
cloud. The
temperature is too high and the gas remains fully ionized even very deep in a
slab of dense gas. An  explorative CLOUDY simulation \citep{CLOUDY} assuming
median values from Table \ref{tbl-2} for luminosity ($\log L_{bol} \approx
47.5$), radius ($\log r \approx$ 17.2), high density and column density ($\log n
= 12$, $\log N = 14$, respectively), indicates that the gas is able to emit just
$\sim 1/50$ of the observed VBC luminosity, with full covering.   The total VBC
emission predicted by CLOUDY of  H$\beta$ is about  2 order of magnitudes lower
than that predicted from simple case B considerations. The amount of gas at the
small distances from the BH derived from the assumption of gravitational
redshift  is not favored as a  significant source of H$\beta$\ emission
for the very luminous HE sources: there is simply not enough gas in the
available volume of radius $\sim 100  \cdot R_{S}$. This conclusion could well
be
of general validity, since VBC shifts of  $\approx 2000$ \kms\ are observed also
in low-luminosity sources.

It is not obvious where some cold gas can survive so close to the
central black hole, at less than $\approx$ 100 Schwarzschild radii.
In the case of an optically thin advection-dominated accretion
flow \citep[ADAF, see e.g.][]{Narayan1996,Narayan1997}, the electron temperature
should be of the order of the virial temperature.  The gas temperature at 100
$R_{S}$\ from a
$10^{9} M_{\odot}$\ black hole is $T_{e}\sim    4 \cdot
10^{6}$\ K, that is too high. On the other hand, the photospheric temperature of
an
geometrically thin, optically thick accretion disk at 100 $R_{S}$\ for
the same mass is $\sim 2 \cdot 10^{3} \dot{M}_{26}^{1/4}$K. The disk is directly
illuminated by only
$\sim 5 \cdot 10^{-4}$ of the total quasar $L$ at 100 $R_{S}$ . Continuum
radiation
could be
also scattered toward the disk (i.e., by an atmosphere of hot electrons). This
case  requires an evaluation of the heating effects of the
continuum radiation over the vertical structure of the disk itself, and 
may be worth a new detailed study, along the line followed by \citet{Dumont90}.

\begin{table*}
\begin{center}
\tiny
\rotate
\caption{Measured quantities for the HE sample: bolometric luminosity, black
hole mass, ${\cal
C}(\frac{1}{4}$),
VBC peak shift, VBC FWHM, fraction of flux in the VBC, the VBC luminosity, as
well as the distances needed for explaining the observed redward displacement
due to gravitational+transverse redshift, as derived from the  ${\cal
C}(\frac{1}{4}$) and from the VBC shift.\label{tbl-2}}
\begin{tabular}{@{}ccrrrrccccccccccl@{}}
\tableline\tableline
Name   &	$L_{bol}(f_{\lambda})$  &  $M_{VP06}$ &
L/$L_{Edd}$ & c(1/4) & VBC shift & FWHM VBC & 
VBC/tot & L($H\beta$) VBC &   $d_{C}$ & $d_{VBC}$\ \\
erg/s & $M_{\odot}$ & & \kms & \kms & \kms & \kms & & erg/s & ($R_{S}$) &
$R_{S}$\\
\tableline
HE0035-2853	&	47.37	&	9.74	&	-0.582	&	1880
&	5470	&	11240	&	0.39	&	1.38E+45	&
119.3	&	41.1	\\
HE0058-3231	&	47.31	&	9.51	&	-0.417	&	2500
&	3350	&	9150	&	0.59	&	2.31E+45	&
90.0	&	67.0	\\
HE0203-4627	&	47.33	&	9.70	&	-0.787	&	1040
&	2060	&	12390	&	0.42	&	7.70E+44	&
217.1	&	109.1	\\
HE0205-3756	&	47.94	&	9.57	&	-0.108	&	2620
&	2270	&	12360	&	0.50	&	5.20E+45	&
85.9	&	99.0	\\
HE0251-5550	&	47.97	&	9.61	&	0.128	&	2890
&	2060	&	11310	&	0.59	&	5.87E+45	&
77.8	&	109.0	\\
HE0349-5249	&	47.72	&	9.73	&	-0.238	&	2200
&	2380	&	11250	&	0.69	&	4.66E+45	&
102.4	&	94.5	\\
HE0436-3709	&	47.20	&	9.56	&	-0.475	&	2050
&	2140	&	11190	&	0.49	&	1.35E+45	&
109.6	&	105.3	\\
HE0926-0201 	&	47.75	&	9.62	&	-0.113	&	800
&	950	&	13010	&	0.56	&	3.74E+45	&
281.1	&	237.1	\\
HE1039-0724	&	47.27	&	10.00	&	-0.942	&	1940
&	2430	&	12170	&	0.35	&	1.13E+45	&
115.6	&	92.5	\\
HE1120+0154	&	47.66	&	9.83	&	-0.434	&	1270
&	2400	&	14040	&	0.53	&	3.24E+45	&
177.4	&	93.6	\\
HE1349+0007	&	47.21	&	9.52	&	-0.404	&	1410
&	3110	&	13070	&	0.44	&	1.18E+45	&
160.0	&	72.3	\\
HE1409+0101	&	47.77	&	10.03	&	-0.796	&	1590
&	2230	&	14560	&	0.54	&	6.16E+45	&
141.4	&	100.7	\\
HE2156-4020	&	47.69	&	9.57	&	-0.152	&	1620
&	1440	&	13050	&	0.60	&	5.96E+45	&
138.9	&	156.7	\\
HE2202-2557 	&	47.28	&	9.65	&	-0.434	&	900
&	3000	&	12340	&	0.48	&	9.63E+44	&
249.8	&	74.9	\\
HE2349-3800 	&	47.37	&	9.45	&	-0.512	&	1100
&	1450	&	11090	&	0.66	&	1.88E+45	&
204.4	&	154.9	\\
HE2355-4621 	&	47.71	&	9.85	&	-0.413	&	1500
&	2150	&	13900	&	0.39	&	4.03E+45	&
149.9	&	104.6	\\\tableline
\end{tabular}
\end{center}
\end{table*}

\section{Could relativistic line profile describe distorted AGN emission line
profiles?}

\begin{figure*}
\begin{center}
 \includegraphics[width=0.95\columnwidth, angle=270]{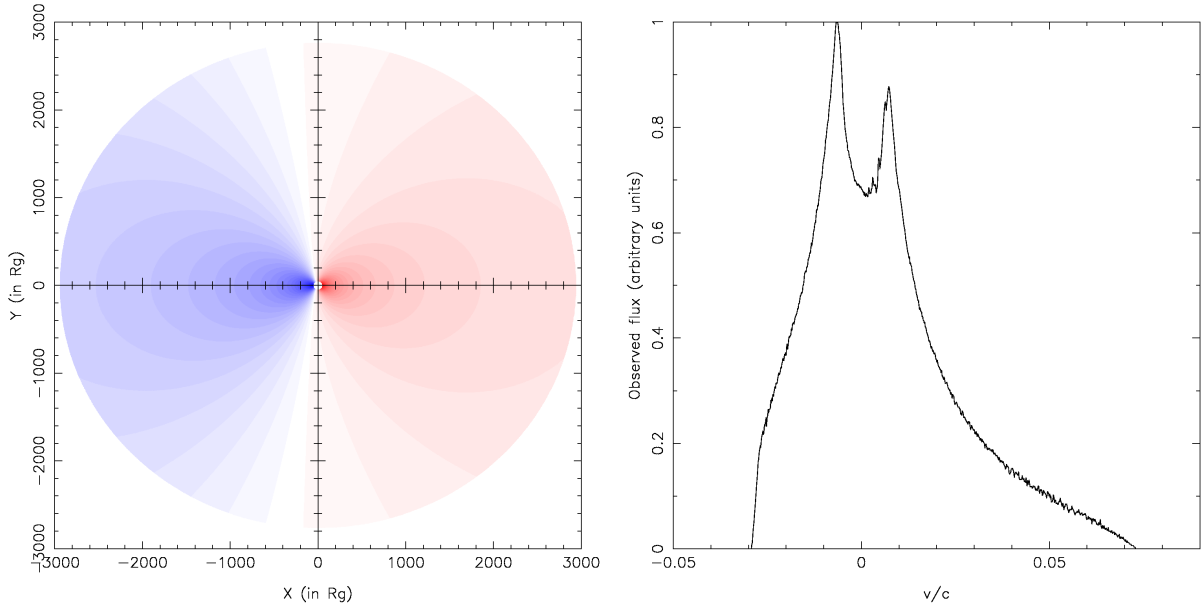}
\caption{Example of an image of a simulated relativistic accretion disk (left
panel) and
corresponding profile of H$\beta$ line (right panel), for inclination
$i=20^{\circ}$, emissivity index of 2.2 and radii Rinn=50 Rg and 
Rout=4000 Rg (where Rg=Rs/2), obtained by ray tracing method in Kerr metric
\citep{Jovanovic2012}.}\label{Disk12}
\end{center}
\end{figure*}

\begin{figure*}
\begin{center}
\includegraphics[width=1\columnwidth, angle=270]{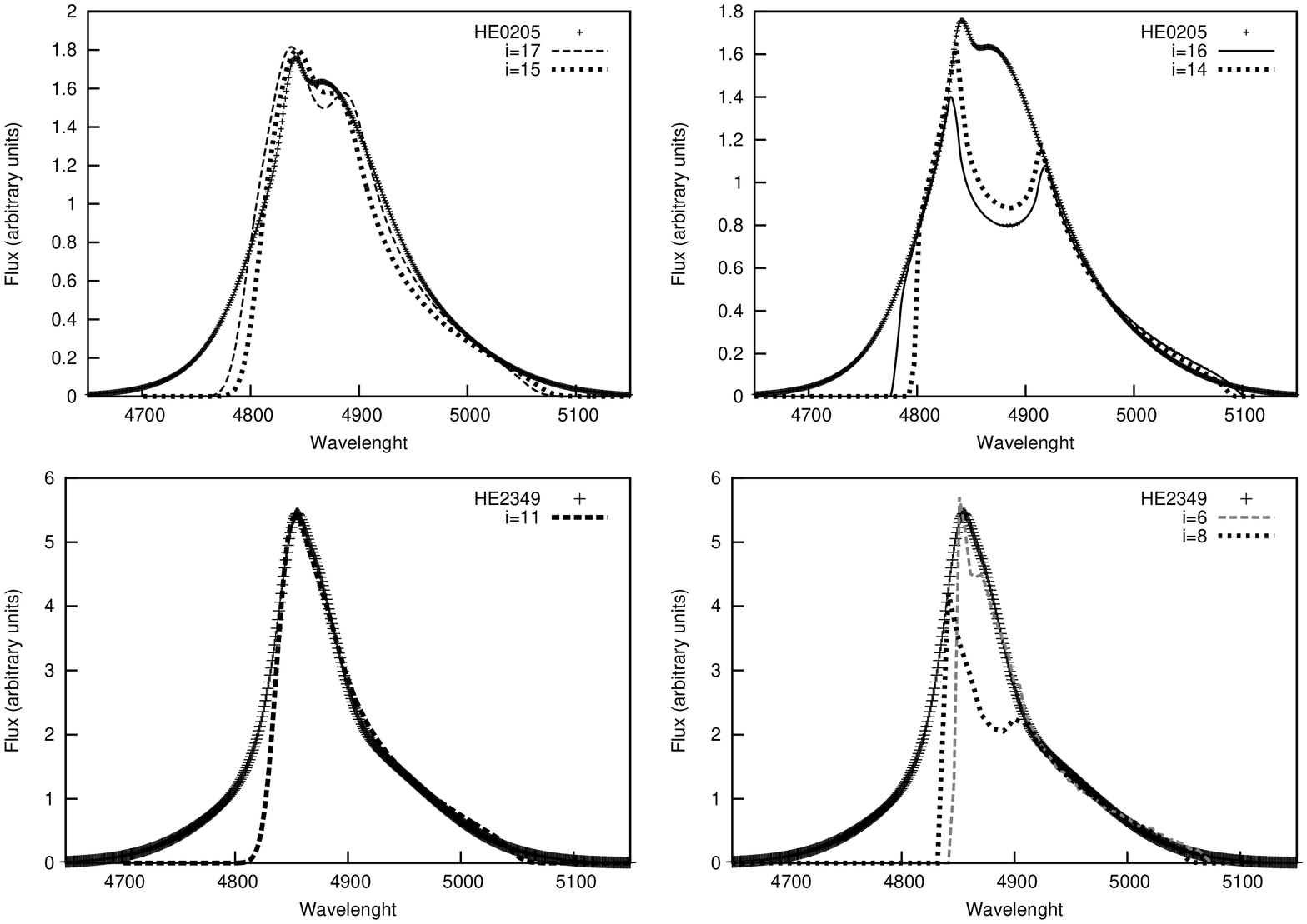}
\caption{Examples of H$\beta$ profiles compared with relativistic disk
profiles calculated using very small inner radius (Rinn $<$ 80 Rg). On the top
panels is the broad component of the H$\beta$ emission line of the object
HE0205, while on the bottom panels are broad emission lines of HE2349. On the
left hand side emission lines are compared with the \citet{CH89} disk model,
while on the right hand panels they are compared with the Kerr
relativistic disk model, obtained by ray tracing method \cite[see
e.g.][]{Cadez98,Jovanovic2012}. The
inclinations are presented on the panels. The other parameters are presented in
the Table \ref{tbl-3}. }\label{Disk}
\end{center}
\end{figure*}

\indent In some AGN, emission lines are detected in X-ray spectral region,
like
Fe K$\alpha$ spectral line (6.4-6.9 keV, depending on ionization state). It
is expected that for high masses ranging from $10^{5}$ to $10^{9}$ solar
masses M$_{\odot}$,
the innermost accretion disk that
surrounds central super-massive black hole should radiate in the X-ray band
\citep{Fabian1989}. Fluorescent /recombination iron K$\alpha$ emission line has
been observed in a number of AGN \cite[see
e.g.][]{Nandra2007,Yaqoob2001,Yaqoob2004,Page2004,Shu2010}. It is thought to be
produced by the hard X-ray power law continuum from the hot corona around
central BH which irradiate relatively cold inner part of accretion disk
\citep{Fabian2000}. Therefore, it is expected to be very broad (velocities 
$\approx$50000 \kms, \citealt{Nandra1997}) and asymmetric due to special
relativistic effects
and large rotational velocities of material emitting near the black hole,
showing very narrow ``blue''
peak and wider and much fainter ``red'' peak \citep{Fabian2000}. General
relativistic effects can further distort the line
shapes, so the red wing becomes more extended towards lower energies for higher
values of the BH spin \citep[see][]{Reynolds2003,Jovanovic2008,Jovanovic2011,
Jovanovic2012},
while the line becomes wider and its red peak brighter.

In comparison to Fe K$\alpha$ emission line, optical emission lines, such as
H$\beta$ line, are expected to be emitted from more distant region then Fe
K$\alpha$, and therefore the gravitational redshift effect in H$\beta$ is
expected to be much weaker.

\indent We simulated relativistic emission line profiles by using ray
tracing relativistic method
\citep{Cadez98,Jovanovic2012,Jovanovic2011,Jovanovic2008, MW06},
assuming that emitting regions are thin equatorial rings in stationary Keplerian
rotation. We simulated the $H\beta$ line profiles using the code which includes 
both special relativistic (transverse Doppler shift and Doppler beaming),
 general relativistic (gravitational redshift) effects on radiation from the 
 accretion disk around SMBH \citep[for more details see e.g.][and 
references therein]{jov09,Jovanovic2012}, and which is based on ray-tracing 
method in the Kerr metric \citep{fan97,Cadez98}. Although this code was
developed for studying the X-ray radiation from innermost regions of the disk,
it can be also applied for simulating the optical emission from its outer
regions \citep[see e.g.][]{Jovanovic2010}.

The example of a
simulated accretion
disk image and the corresponding H$\beta$ line profile are presented in the
Figure \ref{Disk12}.

\indent For the optical emission lines, most commonly used model of the
relativistic
accretion disk is model of \citet{CH89}. It was developed to explain mainly
double peaked profile of optical emission lines, therefore it was mainly tested
on complex emission line profiles of Seyfert type 1.5 AGN (like 3c390.3 and Arp
102B). There were some attempts where \citet{CH89} relativistic disk model was
used to explain single peaked broad emission line shapes
\citep{Pop04,Bon06,Bon09a,Bon09b}. Such
simulations assumed the inner radius of the disk relatively far from the SMBH,
starting from few hundred Rg, where the effects of gravitational redshift were
not significant.\\
\indent Here we tried to simulate broad emission line profiles using 
\citet{CH89} relativistic disk model, assuming that the inner radius of the
disk should originate very close to the SMBH (with values less than hundred
gravitational radii), where the gravitational redshift effect is significant,
with the aim to explain the shape of $H\beta$ extended red wing with
gravitational effects. \\

\indent Simulated profiles are compared to broad, H$\beta$ optical emission
lines from the HE sample, for the cases where centroids of line widths measured
on 1/4 of maximum line intensity were significantly shifted to the red. 
We matched the simulated line
profiles with observed H$\beta$ emission lines in order to see if we could find
the solution which could match their distorted red wing,
assuming the effects of gravitational redshift, considered in
relativistic disk models. \citet{MW06} 
calculated that for the optical lines with resolution of 0.1 \AA\
gravitational redshift in the line profiles becomes important at distances
smaller then 75000 gravitational radius (Rg). 
Since the emissivity weighted radii calculated for HE sample (Table
\ref{tbl-2}) 
are very close to the BH, we assumed that inner radius (Rinn) of the disk model
should start from
similar values. 
The emission lines are compared with the two relativistic accretion disk models:
\citet{CH89} model for relativistic disk in Schwarzschild metric,
and accretion disk model in Kerr metric \citep[see
e.g.][]{Cadez98,Jovanovic2012}.
Starting from the values presented in  Table \ref{tbl-1}  for 
the inner radius of the accretion disk, we searched for the shape that
looked similar as the red side of the emission line profile, while changing
the disk parameters. Examples of simulated disk profiles matched to the
observed broad emission lines of HE0205 and HE2349 are represented in
the Fig. \ref{Disk}, with corresponding parameters of each model presented in 
Table \ref{tbl-3}.

\begin{table}[t]

\scriptsize
\caption{Table lists parameters (inclination ($i$), power law emissivity
index ($e$), inner and outer radius, given in gravitational radii (Rinn, and
Rout), respectively) of the \citet{CH89} disk model (CH89),
and the ray tracing
relativistic disk model \cite[see e.g.][(J12)]{Jovanovic2012}, used to
match H$\beta$ emission line of two spectra from HE sample.
\label{tbl-3}}
\begin{tabular}{@{}llllll}
\tableline
Name    & Disk model & $i$ & $q$ & Rinn (Rg) &
Rout (Rg)\\
\tableline
HE0205  & CH89 &  17 & -2.22 & 80 & 8000 \\
HE0205  & CH89 &  15 & -2.2  & 65 & 12000 \\
HE0205  & J12 & 16 & -2.2 & 75 & 1100      \\
HE0205  &  J12 & 14 & -2.2 & 70 & 1000  \\
HE2349  & CH89 &  11 & -2.5 & 58 & 15000  \\
HE2349  & J12 & 6  & -2.5 & 40 & 5000      \\
HE2349  & J12 & 8  & -2.6 & 53 & 500    \\ 
\tableline
\end{tabular}
 
\end{table}

As could be seen from Fig. \ref{Disk}, the red
wing of the line profile could be well described with 
relativistic accretion disk
model for relatively small range of face on inclinations and inner radius
relatively close to the BH. This is in agreement
with results from \cite{MW06}. 
 
We note here that  such model could not describe the line blue wing. Therefore,
 one would need to assume that, if the accretion disk contributes to some part
of the line flux, it cannot explain the shape of the blue wing. Then at least  
part of the blue wing, and probably part of the line core needs to be emitted
from some other, kinematically different region (outflows, winds, jets, etc.).
Similar models were sugested earlier,  assuming at least two kinematically
different components (the disk with additional emitting component)  relatively
close to BH \citep[see e.g.][]{Corbin97,Pop04,Bon06,Bon09a,Bon09b,Bon2012}. 
 Another difficulty is that the emissivity law ($\sim r^{q}$) is
very steep (with $q$ between -2.2 and -2.6), weighting emission in
favor of the inner disk region, where gas may
be too hot and it is not clear if it can significantly contribute to the total
line emission.

If the red wing of the emission line originate in the accretion disk,
then one could expect to observe in monitoring campaigns of some objects 
that the red wing respond to the continuum variations faster then the blue wing 
\citep[see some cases where this is
observed, e.g.][]{Gaskell13,Done96,Welsh07,Gaskell09}. 
Of course, there are many open questions about accretion disks
\citep{Sulentic99}, 
and many alternative models are suggested to explain the BLR model, like
inflows, outflows, binary or multiple black holes \citep[see
e.g.][]{Bogdanovic15,Bon2012,Popovic12}, and many more.

\section{Summary}

It has been known since 20 years that redward asymmetry of very broad wings are
common in the Type 1 Pop. B AGN \citep[see e.g.][]{Peterson85,MarzSul12}. The
median H$\beta$ profile of Pop. B type of AGN show such asymmetry \citep[see
e.g.][]{MarzSul12}.  This phenomenon could be connected to
the effects of gravitational redshift. Some other alternative explanation of
redshifted emission line could be  that the redshift is due to radial motion,
either inflows \citep{Gaskell13} or outflows \citep[see e.g.][]{ZS90}.

We found that the red wing of the line profile could be well described with the
relativistic 
accretion disk
models for relatively small range of face on inclinations and very
close inner
radii. This indicates that
in some cases the optical emission lines could be distorted by
relativistic effects and gravitational redshift.

However, it is not clear whether there is enough gas close to the central 
black hole that is able to emit the very luminous redshifted emission components
that are observed in the HE quasars. Simple estimates suggest that this might be
unlikely, probably because most gas is expected to be too hot to significantly
emit Balmer lines,  but a detailed model of an illuminated accretion disk is
needed to properly estimate line emissivity. If there is indeed enough
gas to
sustain power law emissivity (with $q$ between $\approx$ -2.2 and $\approx$
-2.6), gravitational redshift could be estimated from the line centroid, and in
this way $z_{grav}$ could be used as a proxy for M$_{grav}$ calculation.

\section{Acknowledgements}

This research is part of the projects 176001 ''Astrophysical
spectroscopy of extragalactic objects'' and 176003 ''Gravitation
and the large scale structure of the Universe'' supported by the Ministry of
Education and Science of the Republic of Serbia.

\end{document}